\documentclass{aa}
\usepackage{natbib}
\bibpunct{(}{)}{;}{a}{}{,} 

\usepackage{graphicx}
\usepackage{rotating}
\usepackage[varg]{txfonts}
\usepackage{longtable,lscape}                         

\newcommand{\hii}{{\rm H{\textsc{ii}}\,}}

\begin{document}

\title{Temperature-averaged and total free-free Gaunt factors for $\kappa$ and Maxwellian distributions of electrons}

   \author{Miguel A. de Avillez\inst{1,2}\and Dieter Breitschwerdt\inst{2}}
   \institute{Department of Mathematics, University of \'Evora, R. Rom\~ao Ramalho 59, 7000 \'Evora, Portugal\\
   \email{mavillez@galaxy.lca.uevora.pt,mavillez@astro.physik.tu-berlin.de}
   \and
    Zentrum f\"ur Astronomie und Astrophysik, Technische Universit\"at Berlin, Hardenbergstrasse 36, D-10623
Berlin, Germany\\
    }

    \date{Received March 16, 2015; accepted May 11, 2015}

   \titlerunning{Temperature averaged free-free Gaunt factors for \emph{Kappa}-distributions}
   \authorrunning{de Avillez \& Breitschwerdt}


\abstract
{}
{Optically thin plasmas may deviate from thermal equilibrium and thus, electrons (and ions) are no longer described by the Maxwellian distribution. Instead they can be described by $\kappa-$distributions. The free-free spectrum and radiative losses depend on the temperature-averaged (over the electrons distribution) and total Gaunt factors, respectively. Thus, there is a need to calculate and make available these factors to be used by any software that deals with plasma emission.}
{We recalculated the free-free Gaunt factor for a wide range of energies and frequencies using hypergeometric functions of complex arguments and the Clenshaw recurrence formula technique combined with approximations whenever the difference between the initial and final electron energies is smaller than $10^{-10}$ in units of $z^{2}Ry$. We used double and quadruple precisions. The temperature-averaged and total Gaunt factors calculations make use of the Gauss-Laguerre integration with 128 nodes.}
{The temperature-averaged and total Gaunt factors depend on the $\kappa$ parameter, which shows increasing deviations (with respect
to the results obtained with the use of the Maxwellian distribution) with decreasing $\kappa$. Tables of these Gaunt factors are provided.}
  {}
\maketitle
%

\section{Introduction}

Astrophysical plasma emission codes are a powerful tool for calculating spectra and energy losses in a plasma such as  the interstellar or intergalactic medium and for comparing the results with observations. However, such
plasmas are complex systems in which frequently made assumptions, like establishing the Maxwell-Boltzmann distribution 
(hereafter referred as Maxwellian distribution) of the electrons and ions, are not always fulfilled and may lead to 
erroneous interpretations of the plasma properties. We have therefore reexamined the nonrelativistic free-free Gaunt 
factor, which is the quantum correction for the semiclassical cross section of \citet{kramers1923}. The factor has been 
the subject of many papers over the years, comprising analytical approximations 
\citep[e.g.,][]{menzel1935, elwert1954, grant1958, brussaard1962, hummer1988, beckert2000} to the exact 
quantum mechanical expressions (in terms of hypergeometric functions) derived, for example, by 
\citet{menzel1935}, \citet[][using nonrelativistic dipole approximation]{sommerfeld1953}, \citet{kulsrud1954}, and 
\citet{biedenharn1956}, and detailed numerical computations first discussed in the seminal work of \citet[][hereafter 
KL61]{karzas1961} and followed by a series of publications during the past five decades.

Karzas \& Latter computed the Gaunt factor using hypergeometric functions of complex variables and presented their 
results in graphical form. The problem reduces to calculating
the solution of differential equations by means of 
power series of the real variable $x$ (which is negative) at two regimes ($\left|x\right|>1$ and $\left|x\right|\leq 1$) 
with a very slow convergence when $\left|x\right|\to 1$. Since KL61, several authors recalculated the Gaunt factor 
with increasing precision and size of parameter space \citep[e.g.,][]{obrien1971, carson1988, hummer1988, 
nicholson1989, janicki1990, sutherland1998, vanhoof2014}. Special care was taken to overcome 
the slow convergence of the solution by (i) redefining the regimes through a change in variables \citep{obrien1971}, 
which has been adopted, in combination with KL61 formulae, in most of the calculations that followed, (ii) increase in 
precision, and (iii) by using the approximation of \citet[][see discussion in \citet{vanhoof2014}, who corrected a term 
in their formulae]{menzel1935}.

The Gaunt factor can be averaged over a distribution of electrons (also known as the temperature-averaged Gaunt 
factor) and then integrated over the full range in frequency (the total Gaunt factor). These in turn are used in the 
determination of the emission spectra and radiative losses by a plasma as a result of this process. In general, the 
averaging is made over a Maxwellian distribution of electrons, thus relying on the  assumption that thermal 
equilibrium is the rule \citep[see, e.g.,][]{karzas1961, gayet1970, armstrong1971, feng1983, carson1988, 
hummer1988, nicholson1989, janicki1990, sutherland1998, vanhoof2014}. 

\begin{figure}[thbp]
        \centering
        \includegraphics[width=0.85\hsize,angle=0]{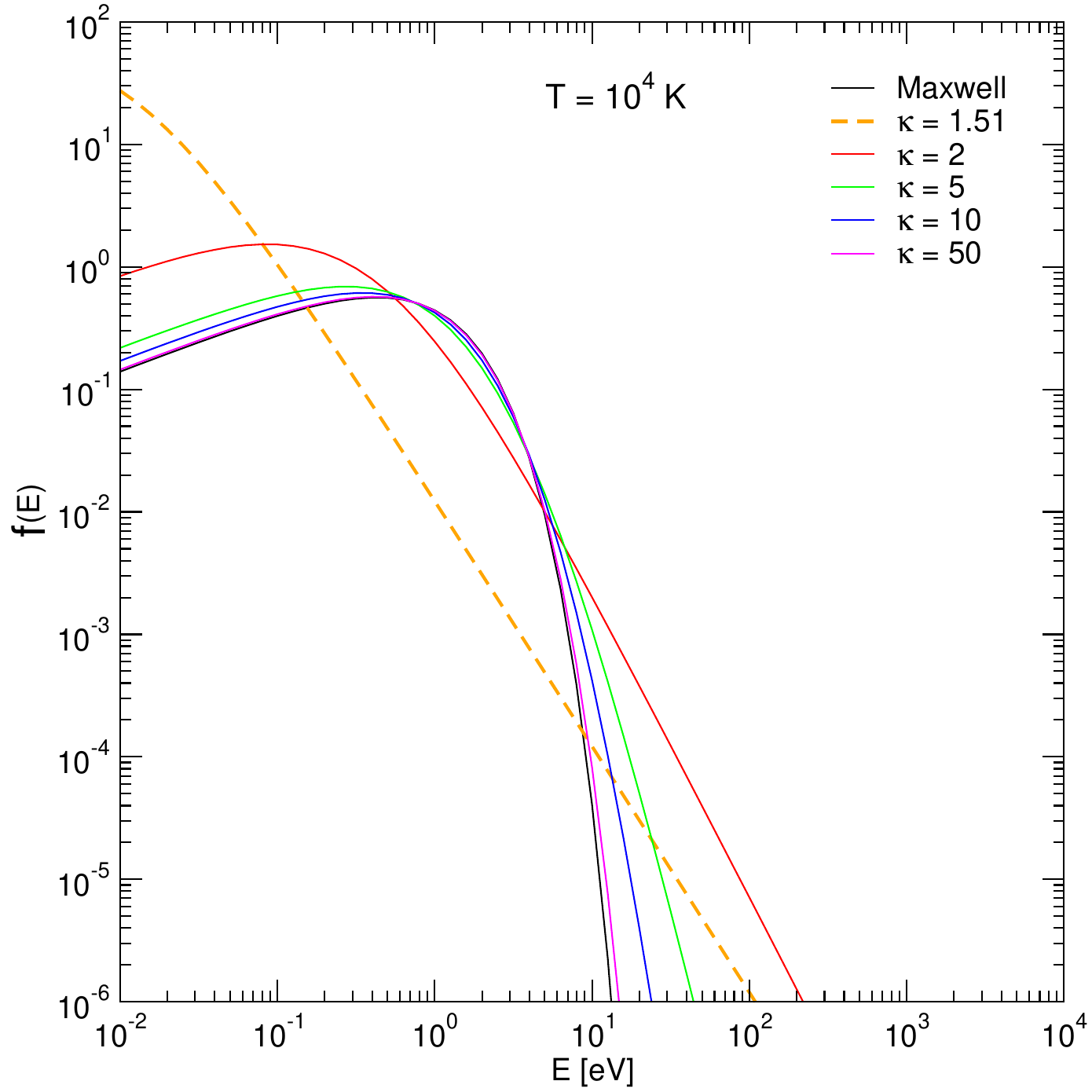}
        
        \includegraphics[width=0.85\hsize,angle=0]{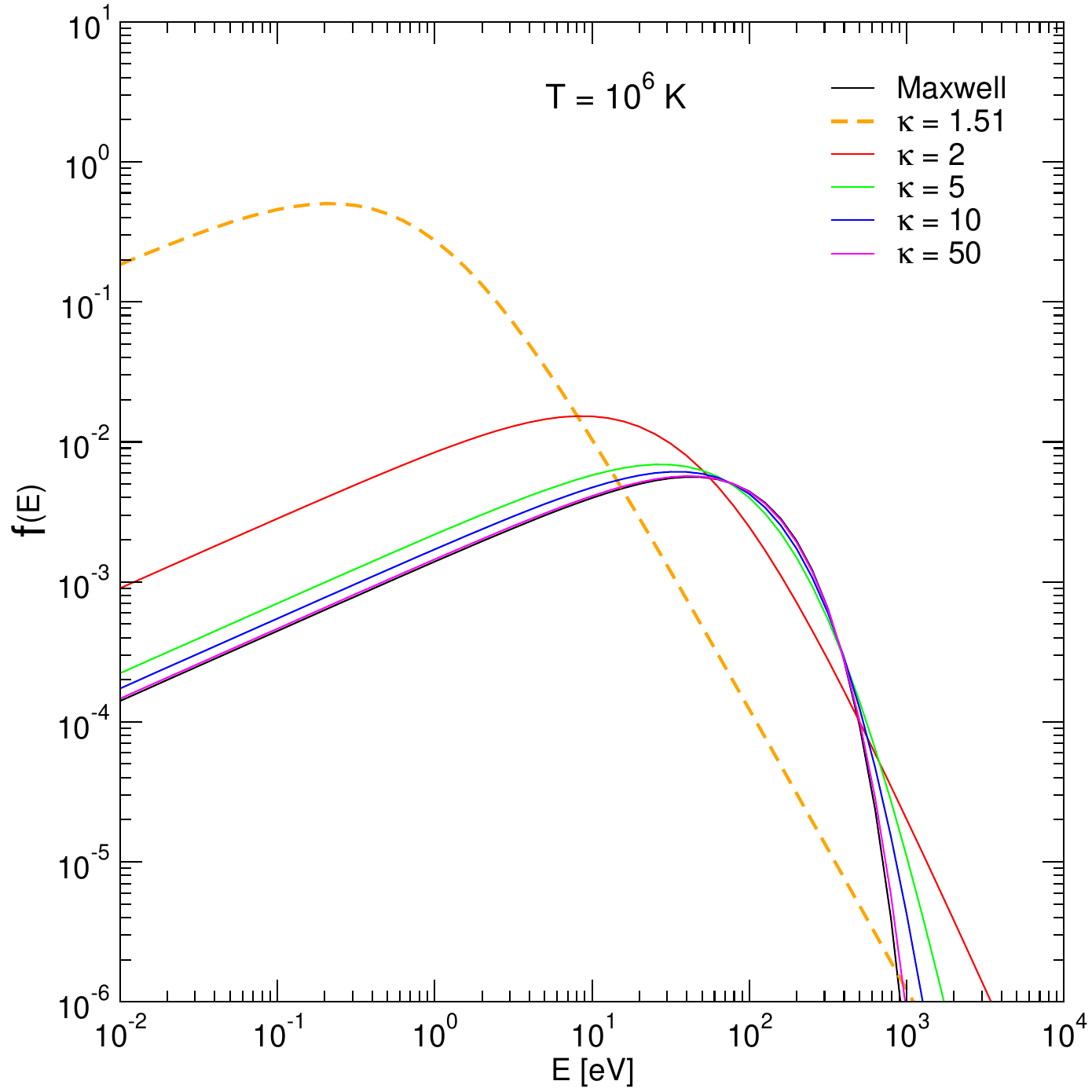}
        \caption{Not normalized Maxwellian and $\kappa$-distributions (with different $\kappa$) at $10^{4}$ K (top) and $10^{6}$ K (bottom panel). The largest deviation of the Maxwellian distribution occurs for the lowest $\kappa,$ showing an increase in electrons at the low- and high-energy ranges when compared to the Maxwellian distribution.}
        \label{kappadist}
\end{figure}

However, this condition may not be attained if high-energy electrons are injected into the system on 
timescales shorter than that needed to achieve thermalization (Livadiotis \& McComas 2009) or when long-range forces 
are present in the plasma \citep{collier2004}. Deviations may also occur in the presence of strong temperature and/or 
density gradients \citep[][see discussion in, e.g., \citet{dudik2011} and references therein]{collier1993, 
collieretal2004}. An example of nonthermal distributions is the so-called $\kappa$-distribution, which was first used 
by \citet{vasyliunas1968} to match the observed electron distribution in the Earth's magnetosphere. $\kappa-$ 
distributions have also been used to explain the discrepancies observed in the abundances and temperatures in \hii 
regions and planetary nebulae when derived using collisional excitation lines and optical recombination lines 
\citep[see][]{binette2012,nicholls2012,nicholls2013}. The $\kappa$-distribution is characterized by a high-energy 
power-law tail and has the 
form
\begin{equation}
f_{\kappa}(E)dE=\frac{2E^{1/2}}{\pi^{1/2}(k_{B}T)^{3/2}} \displaystyle A_{\kappa} \left[ 1+\frac{E}{(\kappa-3/2)k_{B}T}\right]^{-\kappa-1}dE,
\end{equation}
with 
\begin{equation}
A_{\kappa}=\frac{\Gamma(\kappa+1)}{\Gamma(\kappa-1/2)(\kappa-3/2)^{3/2}},
\end{equation}
and $\Gamma(x)$ denoting the gamma function of the variable $x$.  When $\kappa\to \infty,$ the Maxwellian distribution 
is recovered. As $\kappa$ decreases, deviations from the Maxwellian distribution increase, reaching maximum when 
$\kappa$ approaches 3/2 (Fig.~\ref{kappadist}, which displays the $\kappa$ and Maxwellian distributions at 
different temperatures for electrons with energies varying between $10^-2$ and $10^4$ eV). Similarly to the 
Maxwellian distribution, the mean energy of the $\kappa$-distribution is independent of $\kappa$ and is given by 
$\langle E\rangle=3/2k_{B}T$. Hence, $T$ can be defined as the thermodynamic temperature for these distributions. 
For a review of the $\kappa$-distribution and its applications see \citet{pierrard2010} and \citet{livadiotis2009}. 

\citet{dudik2011,dudik2012} calculated the free-free contribution to the emission spectra and to the radiative losses 
of a plasma with the nonthermal distributions ($\kappa$ and $n$ distributions) and evolving under collisional 
ionization equilibrium conditions (CIE), that is, the number of recombinations equals the number of ionizations by 
electron impact. The authors calculated the temperature-averaged Gaunt factors for nonthermal distributions using 
the free-free Gaunt factors of \citet{sutherland1998}. 

The temperature-averaged and total Gaunt factors are needed for detailed simulations involving the coupling of the dynamical and thermal evolutions of the interstellar medium, bubbles and superbubbles (including the Local Bubble), and formation of galaxies, to name just a few. Hence, we carried out detailed calculations of the free-free Gaunt factor and of the temperature-averaged and total Gaunt factors for Maxwellian and $\kappa-$distributions (considering a large grid of $\kappa$ parameters). We present these results in tabulated form to be used in any plasma emission software through convenient one- and two-dimensional interpolations. 

The structure of this paper is as follows: Section 2 describes the calculation of the free-free Gaunt factors. Section 3 deals with the temperature-averaged and total Gaunt factor for Maxwellian and $\kappa$ distributions of electrons. Section 4 describes the tabulated data, and we conclude in Sect. 5 with some final remarks.

\section{Calculations of the Gaunt factor}

The calculation of the free-free Gaunt factor in double and quadruple precision follows the prescription of 
\citet{janicki1990} with some adaptations taken from \citet{carson1988} and \citet{vanhoof2014}. It is assumed that 
an electron with initial energy $E_{i}$ absorbs a photon of energy $h\nu$. Thus, the electron transits to a 
higher state with an energy $E_{f}$=$E_{i}+h\nu$. The free-free Gaunt factor is then given by  (KL61)
\begin{equation}
g_{_{\rm ff}}=
\frac{2\sqrt{3}}{\pi}I_{0}\left[I_{0}\left(\frac{\eta_{i}}{\eta_{f}}+\frac{\eta_{f}}{\eta_{i}}+2\eta_{i}\eta_{f}\right)-2I_{1}\sqrt{1+\eta_{i}^{2}}\sqrt{1+\eta_{f}^{2}}\right],
\end{equation}
where $\eta_{i,f}^{2}=1/\epsilon_{i,f}$, $\epsilon_{i}=E_{i}/Z^{2}Ry$ and $\epsilon_{i}=E_{f}/Z^{2}Ry$ are the 
electron scaled initial and final energies, respectively; $Ry$ is the infinite-mass Rydberg unit of energy, and $Z$ the 
atomic number of the ion.  The functional $I_{l}$ (with $l=0.1$) is defined as
\begin{equation}
I_{l}=\frac{1}{4}\left(\frac{4\eta_{i}\eta_{f}}{(\eta_{i}-\eta_{f})^{2}}\right)^{l+1} e^{\pi\left|\eta_{i}-\eta_{f}\right|/2}\frac{\left|\Gamma(l+1+i\eta_{i})\Gamma(l+1+i\eta_{f})\right|}{\Gamma(2l+2)}G_{l},
\end{equation}
where $G_{l}$ is a real function given by
\begin{equation}
G_{l}=(1-x)^{-i\eta_{i}-i\eta_{f}} {}_{2}F_{1}(a,b;c;x)
\end{equation}
with $x=-4 \eta_{i}\eta_{f}/(\eta_{i}-\eta_{f})^2$ (which is always negative), $a=l+1-i\eta_{f}$, $b=l+1-i\eta_{i}$, and $c=2l+2$; ${}_{2}F_{1}(a,b;c;x)$ is the hypergeometric function, which satisfies the equation
\begin{equation}
x(1-x)F^{\prime\prime}+\left[c-(a+b+1)xF^{\prime}\right]-abF=0.
\end{equation}
$G_{l}$ satisfies the equation (Janicki 1990)
\begin{equation}
x(1-x^{2})G^{\prime\prime}+(1-x)f_{1}\,G^{\prime}+f_{2}\, G=0,
\end{equation}
with $f_{1}=c+x(2d-a-b-1)$ and $f_{2}=x[d^{2}+ab-d(a+b)]-ab+dc]$.
\begin{figure}[thbp]
        \centering
        \includegraphics[width=0.9\hsize,angle=0]{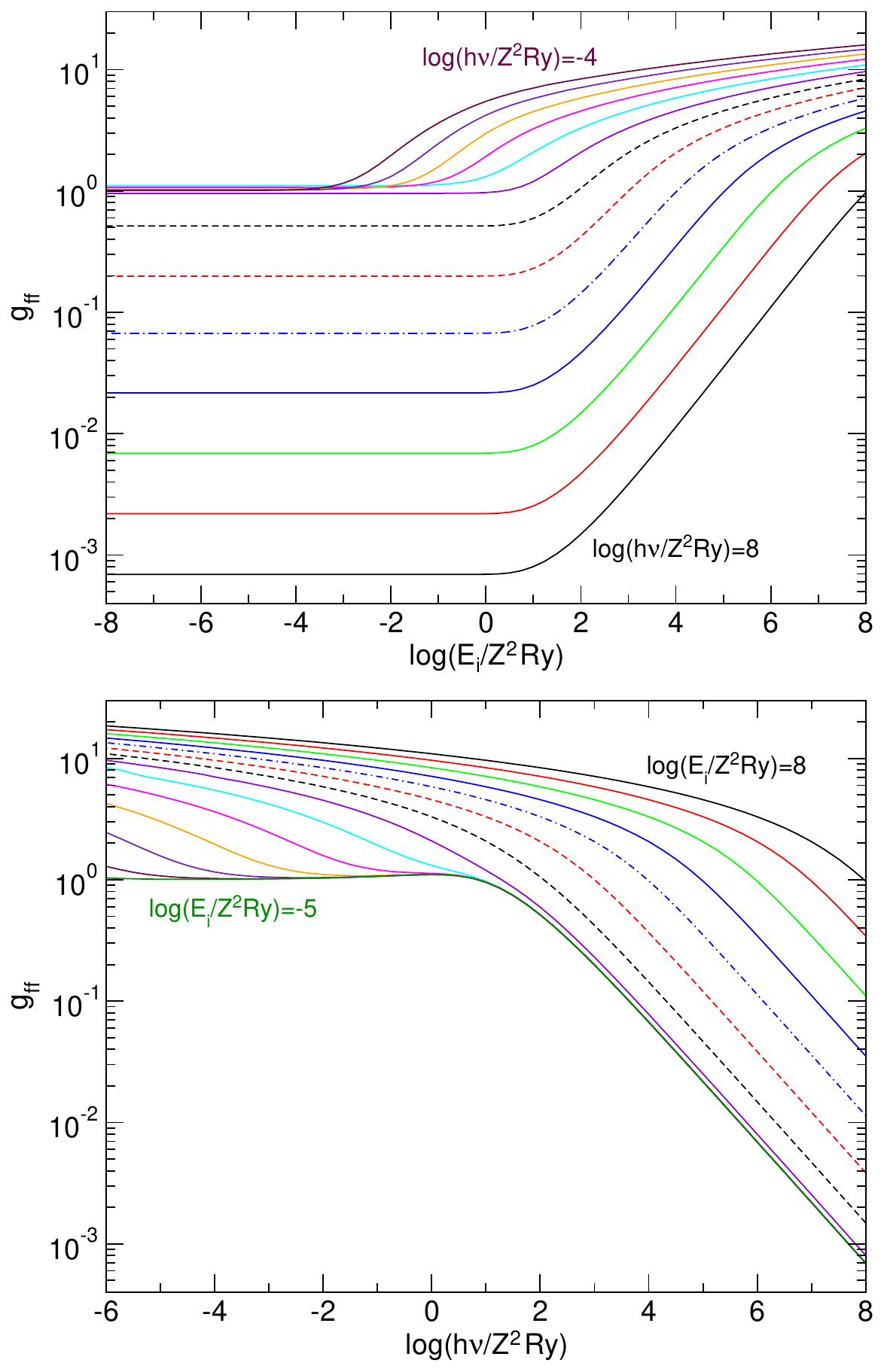}
        \caption{Free-free Gaunt factor variation with the normalized initial electron energy (top panel) for specific photon energies running from $\log h\nu/z^{2}\mbox{Ry}=8$ through $-4$ with steps of dex=1, and with the normalized photon energy (bottom panel) for initial electron energies of $\log E_{i}/z^{2}\mbox{Ry}=-5$ through $8$ with steps of dex=1. }
        \label{gff}
\end{figure}
Thus, the free-free Gaunt factor determination reduces to the calculation of solutions for $G_{l}$, and therefore 
for $I_{l}$, for $l=0,\,1$ in terms of a power series in $x$ for  $\left|x\right|<1$ and in $y=-1/x$ for 
$\left|x\right|>1$ (KL61). These solutions converge very slowly near 1, but a change in variables to \cite{janicki1990}
\begin{equation}
y=\left\{ \begin{array}{lll} 
x/(x-1) & if&\left|x\right|\leq 1.61804 \\
& \\
-1/x & if  & \left|x\right|> 1.61804 
\end{array}\right.
\end{equation}
facilitates the calculation. The threshold results from equating $\left|x\right|/(\left|x\right|-1)$ to $-1/\left|x\right|$ and 
taking into account that $x<0$ (see text above). Another step comprises the usage of the Clenshaw recurrence 
formula to calculate $G_{l}$ at each regime by means of a series of terms. The number of  terms $n$ to be 
considered in the sum to within a certain precision, for example, $10^{-\delta}$, is given by $-\delta/\log \left|x/(x-1)\right|$ 
(for $\left|x\right|\leq 1.61804$) and $-\delta/\log \left|-1/x\right|$ ($\left|x\right|> 1.61804$). For further details see 
\citet{janicki1990}.

In the range $\epsilon_{i,f}^{3/2}/w\leq 10^{-4}$, with $w=\epsilon_{f}-\epsilon_{i}=h\nu/Z^{2}Ry$, the exact 
solution fails \citep[see][]{carson1988, hummer1988, vanhoof2014}. Hence, the approximation of \citet{menzel1935} 
with the correction by \citet{vanhoof2014} is used:
\begin{equation}
g_{_{\rm ff}}=1+c_{1}\frac{1+k^{2}}{\chi^{2}}+c_{2}\frac{1-4/3k^{2}+k^{4}}{\chi^{4}}+c_{3}\frac{1-1/3k^{2}-1/3k^{4}+k^{6}}{\chi^{6}}
,\end{equation}
with $k=\eta_{f}/\eta_{i}$, $\chi=\left[(1-k^{2})\eta_{f}\right]^{1/3}$, $c_{1}=0.1728260369...$, 
$c_{2}=-0.04959570168...$, and $c_{3}=-0.01714285714...$. The maximum error in this approximation is smaller 
than $5.5\times 10^{-10}$ \citep{vanhoof2014}.

Figure~\ref{gff} displays the variation of the free-free Gaunt factor with the normalized initial electron energy (top 
panel) for specific photon energies running from $\log h\nu/z^{2}\mbox{Ry}=8$ through $-4$ with steps of dex=1, 
and with the normalized photon energy (bottom panel) for initial electron energies of $\log E_{i}/z^{2}\mbox{Ry}=-5$ 
through $8$ with steps of dex=1. These Gaunt factors overlap with those reported by \citet{vanhoof2014} and 
\citet{sutherland1998}.

\section{Temperature-averaged and total Gaunt factor}

The energy spectrum by free-free emission from electrons with an energy distribution $f(E)$ is given by \citep[see, 
e.g.,][]{kwok2007}
\begin{equation}
\label{spectrum}
\frac{dP_{_{\rm ff}}}{d \nu}=\frac{8 \pi^{2}}{c^{3}}\left(\frac{2}{3 m_{_{e}}}\right)^{3/2} e^{6} z^{2}n_{e} 
n_{Z,z}\displaystyle \int_{h\nu}^{+\infty}\frac{1}{E^{1/2}}f(E) g_{_{\rm ff}}(E,\nu) dE
,\end{equation}
where $\nu$ is the frequency of the emitted photon, $T$ is the temperature, $k_{B}$ is the Boltzmann constant, 
$n_{e}$ is the electron density, $n_{_{Z,z}}$ is the number density of ion with atomic number $Z$ and ionization 
stage $z$. For electrons with a $\kappa$ or Maxwellian ($\kappa\to\infty$) distribution, and after suitable 
change of variables, Eq. (\ref{spectrum}) becomes 
\begin{eqnarray}
\label{power}
\frac{dP_{_{\rm ff}}}{du} &=& C_{_{\rm ff}} z^{2} n_{e} n_{_{Z,z}} T^{1/2} \times  \\ 
                   & \times & \left\{ \begin{array}{lllc} 
        \displaystyle  \int_{0}^{+\infty}g_{_{\rm ff}}(\gamma^{2},u)\frac{A_{\kappa}}{\left[1+\frac{x+u}{\kappa-3/2}\right]^{\kappa+1}}dx & if & \kappa> 3/2 \\
                && \nonumber\\
                \displaystyle  e^{-u} \int_{0}^{+\infty}g_{_{\rm ff}}(\gamma^{2},u)e^{-x}dx & if & \kappa\to \infty, 
        \end{array}\right.
\end{eqnarray}
where $\displaystyle C_{_{\rm ff}}=16 \left(\frac{2 \pi}{3 m_{_{e}}}\right)^{3/2} \frac{e^{6} k_{B}^{1/2}}{hc^{3}}=1.4256\times 10^{-27}$, and the parameters $x$, $u$ and $\gamma$ have the forms
\begin{equation}
x=\frac{E}{k_{B}T}-\frac{h\nu}{k_{B}T},\, u=\frac{h\nu}{k_{B}T} \mbox{~~and~~} 
\gamma^{2}=\frac{z^{2}Ry}{k_{B}T}=z^{2}\frac{1.579\times 10^{5} \mbox{K}}{T}.
\end{equation}
The integral on the right-hand side of Eq. (\ref{power}) is the temperature-averaged Gaunt factor (KL61). Integration of \ Eq.
(\ref{power}) over the photon frequency spectrum gives the total free-free power associated with an ion $(Z,z)$
\begin{equation}
P_{_{\rm ff}}(T)=C_{_{\rm ff}} z^{2} n_{e}n_{_{Z,z}}T^{1/2} \int_{0}^{+\infty}  \langle g_{_{\rm 
ff}}(\gamma^{2},u)\rangle f(u) du
\label{cooling}
,\end{equation}
with $f(u)=e^{-u}$ (for $\kappa \to \infty$; Maxwellian distribution) and $f(u)=1$ for $\kappa>3/2$ ($\kappa$ 
distribution). The total free-free Gaunt factor is defined as (see, e.g., KL61)
\begin{equation}
g_{_{\rm ff}}(T)=\int_{0}^{+\infty}\langle g_{_{\rm ff}}(\gamma^{2},u) \rangle e^{-u} du
,\end{equation}
and, thus, can be calculated for both distributions.

\begin{figure}[thbp]
\centering
\includegraphics[width=0.85\hsize,angle=0]{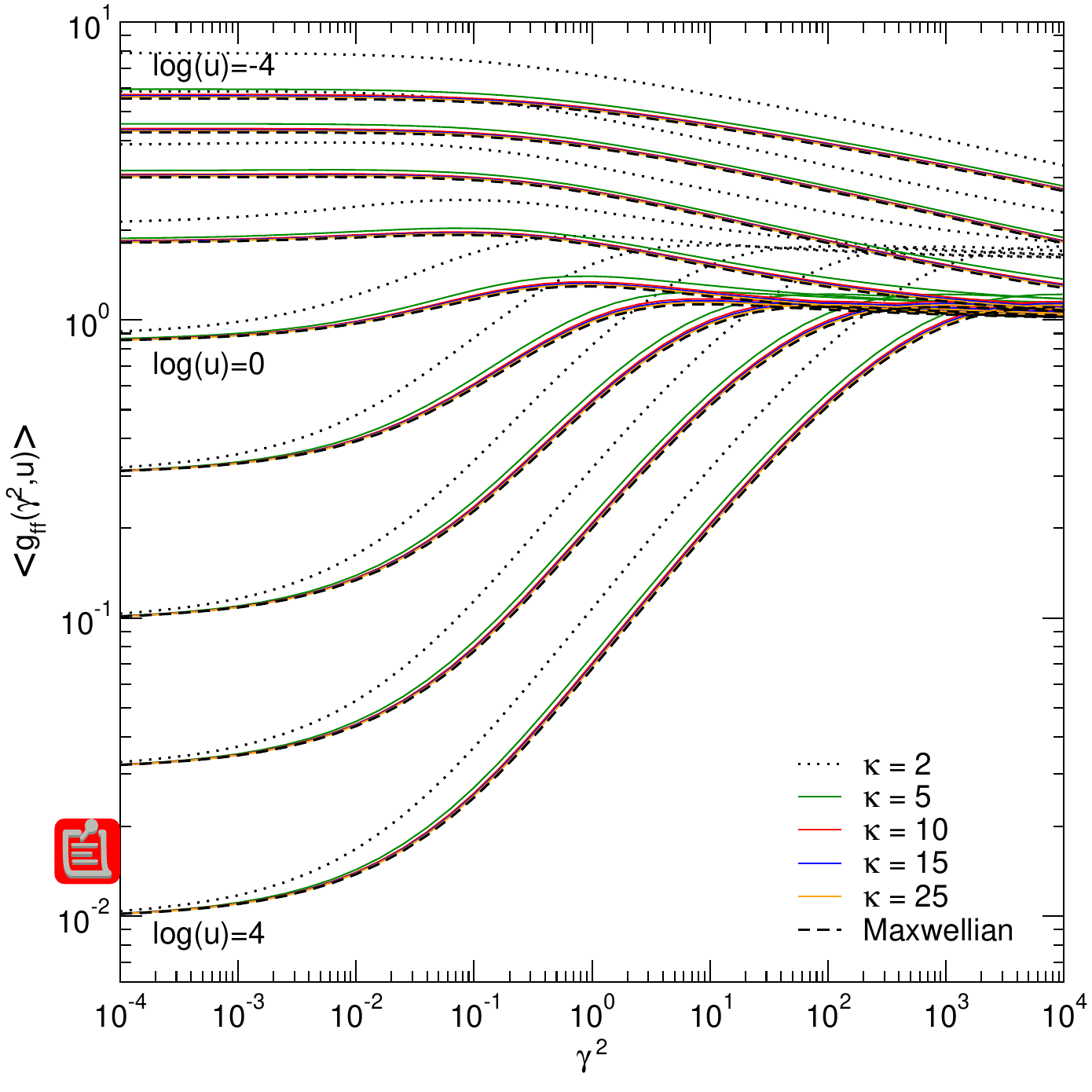}
\includegraphics[width=0.85\hsize,angle=0]{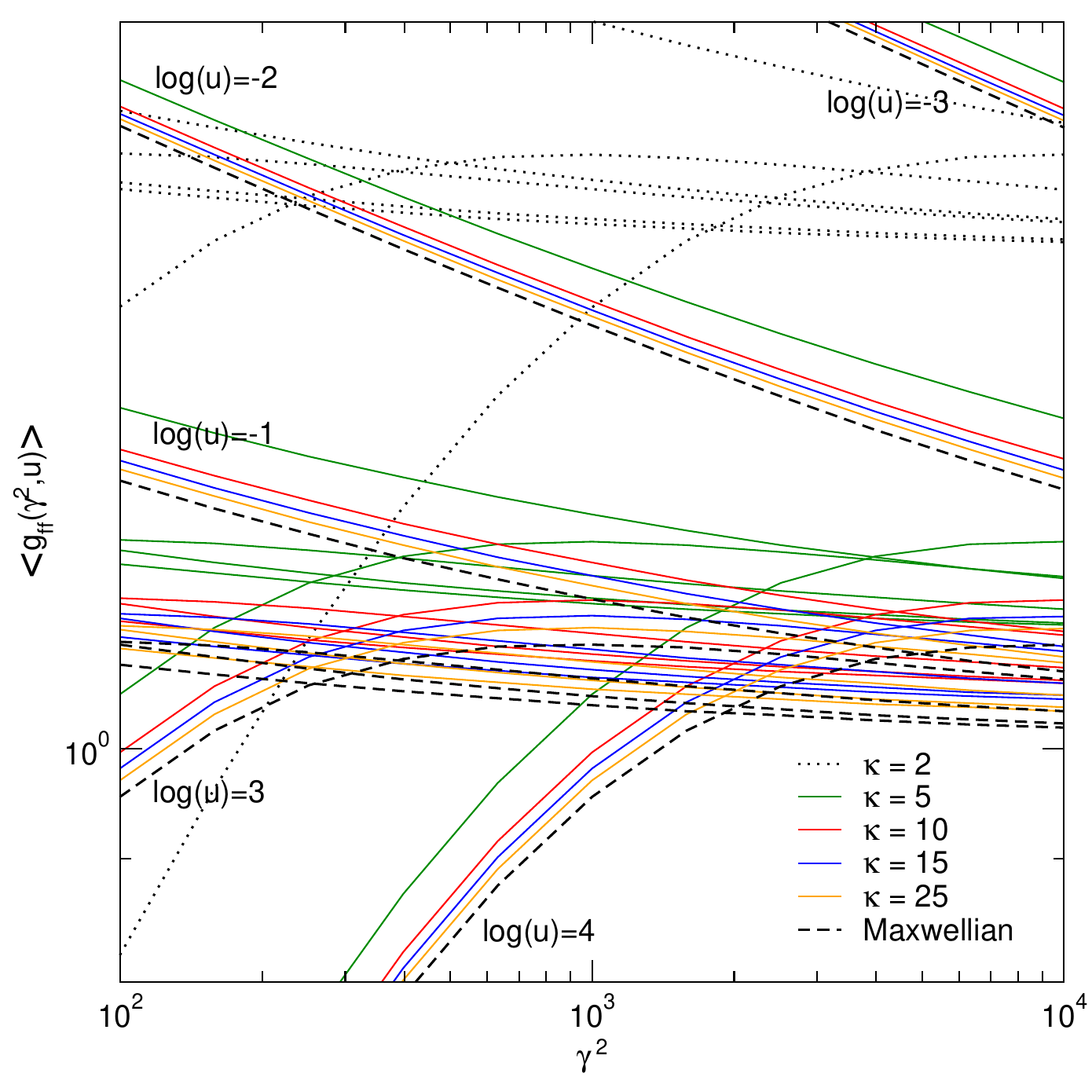}
\caption{Temperature-averaged Gaunt factors calculated for $\kappa=2$, 5, 10, 15, and 25) and Maxwellian 
distributions of electrons for the range $10^{-4}\leq \gamma^{2}\leq10^{4}$ (top panel) and zoomed-in to the region 
$\gamma^{2}\in[10^{2},10^{4}]$ and $\langle g_{_{ff}}(\gamma^{2},u)\rangle \in[0.8,2]$ (bottom panel).}
\label{temperature_averaged}
\end{figure}
\begin{figure}[thbp]
\centering
\includegraphics[width=0.9\hsize,angle=0]{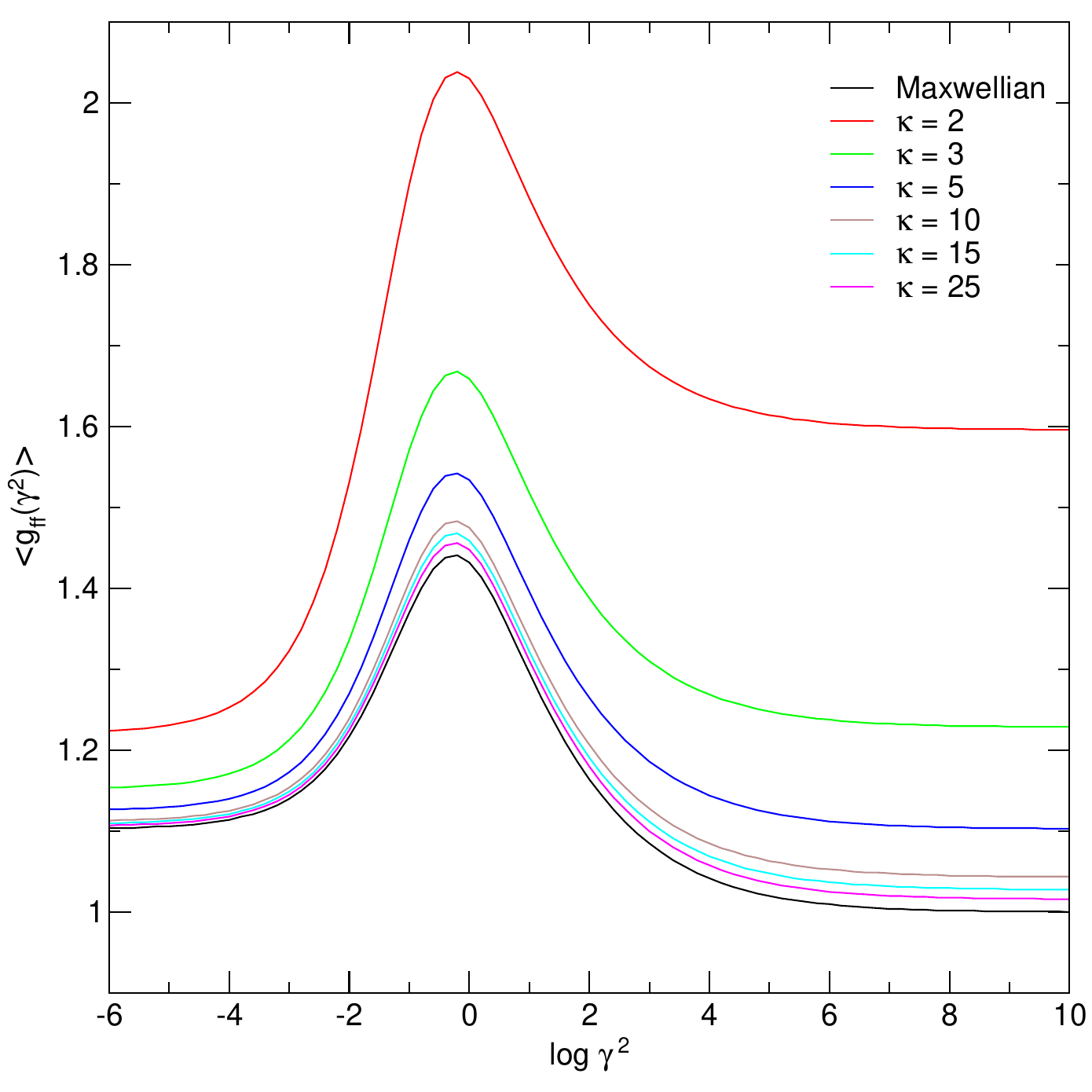}
\caption{Total free-free Gaunt factor calculated for $\kappa$ (2, 3, 5, 10, 15, and 25) and Maxwellian distributions. 
The maximum of $g_{\rm ff}(T)$ moves to the left with the increase in $\kappa$.\label{total_ff}}
\end{figure}

Figures~\ref{temperature_averaged} and~\ref{total_ff} display the variation with $\gamma^{2}$ of the temperature-averaged, $\langle g_{_{\rm ff}}(\gamma^{2},u)\rangle$, and total free-free, $g_{_{ff}}(T)$ Gaunt factor. The results correspond to $\kappa=2$, 5, 10, 15, 25 and $\infty$ (the latter value corresponds to the Maxwellian distribution). 

The temperature-averaged Gaunt factors depend on the $\kappa$ parameter - with the decrease of $\kappa,$ the 
deviations of $\langle g_{_{\rm ff}}(\gamma^{2},u)\rangle$ increase with respect to the Maxwellian-averaged value 
(Fig.~\ref{temperature_averaged}). The strongest deviation always occurs for $\kappa\to 3/2$. These deviations 
depend strongly on the $\gamma^{2}$ (i.e., on the temperature) and mildly on the $u$ (i.e., on the frequency) 
parameters. As $\gamma^{2}$ tends to $-\infty,$ the $\langle g_{_{\rm ff}}(\gamma^{2},u)\rangle$ for different 
$\kappa$ approximate the Maxwellian-averaged value (black dashed line; top panel of 
Fig.~\ref{temperature_averaged}), depending on the value of $u$. With the decrease in $u,$ the blending occurs for 
lower $\gamma^{2}$ than for $u=5$. With an increase in $\gamma^{2}$ (decrease in temperature), the deviations 
become larger and more pronounced for $\gamma^{2}>10$ (bottom panel of Fig.~\ref{temperature_averaged}). 
However, regardless of the value of $\gamma^{2}$ , the $\langle g_{_{\rm ff}}(\gamma^{2},u)\rangle$ shows a small 
variation for all $\kappa\geq 5$ even for $\gamma^{2}>10$.

The total Gaunt factor shows a similar dependence on $\kappa$ for the different $\gamma^{2}$ as $\langle g_{_{\rm 
ff}}(\gamma^{2},u)\rangle$. That is, the deviations increase with increasing $\gamma^{2}$ (Fig.~\ref{total_ff}). The 
total Gaunt factor maximum decreases with increasing $\kappa$ from 2.038 to 1.441 for $\kappa=2$ and 
$\kappa\to \infty$, respectively. When $\gamma^{2}\to \infty,$ the total Gaunt factor tends to 1.596 for $\kappa=2,$ 
decreasing toward 1.0 with $\kappa\to \infty$ (the Maxwellian distribution). With $\gamma^{2}\to -\infty$ $g_{_{\rm 
ff}}(T)\to 1.224$  and 1.1 for $\kappa=2$ and $\kappa \to \infty $, respectively.
As the total Gaunt factor decreases with increase in $\kappa,$ the losses of energy due to free-free emission follow the same path as a result of Eq. (\ref{cooling}).

\section{Tables}

Temperature-averaged and total free-free Gaunt factors calculated for $\kappa$ (2, 3, 5, 10, 15, 25, and 50) and 
Maxwellian distributions are displayed in Tables A.1-A.8 ($\langle g_{_{\rm ff}}(\gamma^{2},u)\rangle$ vs. 
$\gamma^{2}$ for different $u$) and B.1 (appendices A and B, respectively). The parameter space in display 
comprises $\gamma^{2}\in[10^{-8},10^{10}]$ and $u\in [10^{-4},10^{4}]$, but our calculations, and the publically 
available data in http://www.lca.uevora.pt, cover a wider range in these parameters. More data can be provided by 
the authors upon request.

\section{Final remarks}

Optically thin plasmas in the interstellar medium may deviate from thermal equilibrium and thus, electrons are no 
longer described by the Maxwellian distribution. Instead they can be described by $\kappa-$distributions. These 
have been used to explain the deviations between derived abundances and temperatures in \hii regions and planetary 
nebulae. Free-free emission dominates the cooling function of optically thin plasmas at temperatures greater than 
$10^7$ K. The free-free spectrum and radiative losses depend on the temperature-averaged (over the electron 
energy) and total Gaunt factors, respectively. Thus, there is a need to calculate and make available these factors to 
be used by any software that deals with plasma emission. 

Notable astrophysical plasmas, which are dominated by free-free emission, apart from supernova remnants and 
superbubbles in the interstellar medium, are the  intracluster and intergalactic media in clusters of galaxies, 
for instance, where the hot medium dominates the baryonic matter. In particular, merger events in which smaller 
clusters and groups of galaxies fall into larger ones are accompanied by shocks, and hence deviations from thermal 
equilibrium are expected. On a larger scale still, structure formation shocks can develop as a consequence of gas 
infall onto dark matter halos, thereby converting gravitational into thermal energy \citep[see, e.g.,][]{pfrommer2006}. 
The missing-baryon problem and its possible solution by the existence of a widespread warm hot intergalactic 
medium \citep[see][]{cen1999, cen2006} is another example for the importance of free-free emission at high 
temperatures. In many of these contexts, $\kappa$-distributions are therefore expected to provide a better description 
than assuming a Maxwellian.   

Here we have recalculated the nonrelativistic free-free Gaunt factor and its temperature averaged over a large 
spectrum of $\kappa$ parameters (including the Maxwellian distribution) and integrated it over the frequency to 
obtain the total Gaunt factor. We found that the $\kappa$ parameter most affects the temperature-averaged and total 
Gaunt factor at lower temperatures. 

\begin{acknowledgements}
The authors thank the anonymous referee for the comments improving the paper. This research was supported by 
the project "Hybrid computing using accelerators \& coprocessors-modelling nature with a novell approach" (PI: 
M.A.) funded by the InAlentejo program, CCDRA, Portugal. Partial support to M.A. and D.B. was provided by the 
\emph{Deut\-sche For\-schungs\-ge\-mein\-schaft}, DFG project ISM-SPP 1573. The computations made use of the 
ISM Xeon Phi Cluster of the Computational Astrophysics Group, University of \'Evora.
\end{acknowledgements}

\bibliography{bibliography} 

\newpage
\begin{appendix}
\section{Tables of the temperature-averaged free-free Gaunt factor for different $\kappa$}

\begin{table*}
        \centering
        \caption{Temperature-averaged free-free Gaunt factor vs. $\gamma^{2}$ for different $u$ and Maxwellian 
        distribution.}
\begin{tabular}{rccccccccc}
                \hline\hline
                \noalign{\smallskip}
                log $\gamma^{2}$   & \multicolumn{9}{c}{log $u$} \\
                & -4.00 &       -3.00 &       -2.00 &       -1.00 &        0.00 &        1.00 &        2.00 &        3.00 &        4.00 \\    
                \hline
                \noalign{\smallskip}
 -8.00 &   5.528E+00 &   4.259E+00 &   3.002E+00 &   1.806E+00 &   8.424E-01 &   3.035E-01 &   9.801E-02 &   
 3.107E-02 &  9.826E-03 \\
 -7.80 &   5.528E+00 &   4.259E+00 &   3.002E+00 &   1.806E+00 &   8.425E-01 &   3.035E-01 &   9.802E-02 &   
 3.107E-02 &  9.827E-03 \\
 -7.60 &   5.528E+00 &   4.259E+00 &   3.002E+00 &   1.806E+00 &   8.425E-01 &   3.035E-01 &   9.803E-02 &   
 3.107E-02 &  9.828E-03 \\
 -7.40 &   5.528E+00 &   4.259E+00 &   3.002E+00 &   1.807E+00 &   8.426E-01 &   3.035E-01 &   9.804E-02 &   
 3.108E-02 &  9.830E-03 \\
 -7.20 &   5.528E+00 &   4.259E+00 &   3.002E+00 &   1.807E+00 &   8.426E-01 &   3.036E-01 &   9.806E-02 &   
 3.108E-02 &  9.831E-03 \\
 -7.00 &   5.528E+00 &   4.259E+00 &   3.002E+00 &   1.807E+00 &   8.427E-01 &   3.036E-01 &   9.808E-02 &   
 3.109E-02 &  9.834E-03 \\
\hline
\end{tabular}
\end{table*}

\begin{table*}
        \centering
\begin{tabular}{rccccccccc}
        \multicolumn{10}{l}{{\bf{\tablename} \thetable{}} -- Continuation.}\\[0.5ex]
        \hline\hline\\ [-2ex]
        \noalign{\smallskip}
        log $\gamma^{2}$   & \multicolumn{9}{c}{log $u$} \\
        & -4.00 &       -3.00 &       -2.00 &       -1.00 &        0.00 &        1.00 &        2.00 &        3.00 &        4.00 \\    
        \hline
        \noalign{\smallskip}
  4.00 &   2.703E+00 &   1.808E+00 &   1.280E+00 &   1.076E+00 &   1.024E+00 &   1.020E+00 &   1.036E+00 &   
  1.068E+00 &  1.104E+00 \\
  4.20 &   2.600E+00 &   1.737E+00 &   1.247E+00 &   1.065E+00 &   1.021E+00 &   1.017E+00 &   1.031E+00 &   
  1.060E+00 &  1.100E+00 \\
  4.40 &   2.500E+00 &   1.671E+00 &   1.218E+00 &   1.056E+00 &   1.018E+00 &   1.015E+00 &   1.027E+00 &   
  1.053E+00 &  1.093E+00 \\
  4.60 &   2.402E+00 &   1.608E+00 &   1.191E+00 &   1.049E+00 &   1.015E+00 &   1.013E+00 &   1.023E+00 &   
  1.046E+00 &  1.085E+00 \\
\hline
\end{tabular}
\end{table*}

\begin{table*}
        \centering
        \caption{Temperature-averaged free-free Gaunt factor vs. $\gamma^{2}$ for different $u$ and 
        $\kappa-$distribution: $\kappa=2$.}
        \begin{tabular}{rccccccccc}
                \hline\hline
                \noalign{\smallskip}
                log $\gamma^{2}$   & \multicolumn{9}{c}{log $u$} \\
                & -4.00 &       -3.00 &       -2.00 &       -1.00 &        0.00 &        1.00 &        2.00 &        3.00 &        4.00 \\    
                \hline
                \noalign{\smallskip}
 -8.00 &   7.857E+00 &   5.836E+00 &   3.859E+00 &   2.101E+00 &   8.838E-01 &   3.054E-01 &   9.814E-02 &   
 3.110E-02 &  9.836E-03 \\
 -7.80 &   7.857E+00 &   5.836E+00 &   3.860E+00 &   2.101E+00 &   8.839E-01 &   3.055E-01 &   9.816E-02 &   
 3.111E-02 &  9.839E-03 \\
 -7.60 &   7.857E+00 &   5.836E+00 &   3.860E+00 &   2.101E+00 &   8.840E-01 &   3.055E-01 &   9.818E-02 &   
 3.111E-02 &  9.841E-03 \\
 -7.40 &   7.857E+00 &   5.836E+00 &   3.860E+00 &   2.101E+00 &   8.841E-01 &   3.056E-01 &   9.820E-02 &   
 3.112E-02 &  9.843E-03 \\
 -7.20 &   7.857E+00 &   5.837E+00 &   3.860E+00 &   2.101E+00 &   8.843E-01 &   3.056E-01 &   9.823E-02 &   
 3.113E-02 &  9.846E-03 \\
 -7.00 &   7.857E+00 &   5.837E+00 &   3.860E+00 &   2.101E+00 &   8.845E-01 &   3.057E-01 &   9.826E-02 &   
 3.114E-02 &  9.849E-03 \\
\hline
                \end{tabular}
        \end{table*}
        
        \begin{table*}
                \centering

\begin{tabular}{rccccccccc}
        \multicolumn{10}{l}{{\bf{\tablename} \thetable{}} -- Continuation.}\\[0.5ex]
        \hline\hline\\ [-2ex]
        \noalign{\smallskip}
        log $\gamma^{2}$   & \multicolumn{9}{c}{log $u$} \\
        & -4.00 &       -3.00 &       -2.00 &       -1.00 &        0.00 &        1.00 &        2.00 &        3.00 &        4.00 \\    
        \hline
        \noalign{\smallskip}
  4.00 &   3.294E+00 &   2.291E+00 &   1.816E+00 &   1.657E+00 &   1.621E+00 &   1.625E+00 &   1.652E+00 &   
  1.704E+00 &  1.762E+00 \\
  4.20 &   3.168E+00 &   2.222E+00 &   1.789E+00 &   1.649E+00 &   1.618E+00 &   1.621E+00 &   1.644E+00 &   
  1.691E+00 &  1.756E+00 \\
  4.40 &   3.047E+00 &   2.158E+00 &   1.765E+00 &   1.642E+00 &   1.614E+00 &   1.618E+00 &   1.638E+00 &   
  1.680E+00 &  1.745E+00 \\
  4.60 &   2.933E+00 &   2.100E+00 &   1.744E+00 &   1.635E+00 &   1.612E+00 &   1.615E+00 &   1.632E+00 &   
  1.669E+00 &  1.731E+00 \\
\hline
        \end{tabular}
\end{table*}

\begin{table*}
        \centering
        \caption{Temperature-averaged free-free Gaunt factor vs. $\gamma^{2}$ for different $u$ and 
        $\kappa-$distribution: $\kappa=3$.}
        \begin{tabular}{rccccccccc}
                \hline\hline
                \noalign{\smallskip}
                log $\gamma^{2}$   & \multicolumn{9}{c}{log $u$} \\
        & -4.00 &       -3.00 &       -2.00 &       -1.00 &        0.00 &        1.00 &        2.00 &        3.00 &        4.00 \\    
        \hline
        \noalign{\smallskip}
         -8.00 &   6.444E+00 &   4.888E+00 &   3.351E+00 &   1.930E+00 &   8.594E-01 &   3.041E-01 &   9.806E-02 &   
         3.108E-02 &  9.830E-03 \\
         -7.80 &   6.444E+00 &   4.888E+00 &   3.351E+00 &   1.930E+00 &   8.594E-01 &   3.041E-01 &   9.807E-02 &   
         3.108E-02 &  9.831E-03 \\
         -7.60 &   6.444E+00 &   4.888E+00 &   3.351E+00 &   1.930E+00 &   8.595E-01 &   3.041E-01 &   9.808E-02 &   
         3.109E-02 &  9.833E-03 \\
         -7.40 &   6.444E+00 &   4.888E+00 &   3.352E+00 &   1.931E+00 &   8.596E-01 &   3.042E-01 &   9.809E-02 &   
         3.109E-02 &  9.835E-03 \\
         -7.20 &   6.444E+00 &   4.888E+00 &   3.352E+00 &   1.931E+00 &   8.597E-01 &   3.042E-01 &   9.812E-02 &   
         3.110E-02 &  9.837E-03 \\
         -7.00 &   6.444E+00 &   4.888E+00 &   3.352E+00 &   1.931E+00 &   8.598E-01 &   3.043E-01 &   9.814E-02 &   
         3.111E-02 &  9.840E-03 \\
\hline
                        \end{tabular}
                \end{table*}
                
                \begin{table*}
                        \centering 
\begin{tabular}{rccccccccc}
        \multicolumn{10}{l}{{\bf{\tablename} \thetable{}} -- Continuation.}\\[0.5ex]
        \hline\hline\\ [-2ex]
        \noalign{\smallskip}
        log $\gamma^{2}$   & \multicolumn{9}{c}{log $u$} \\
        & -4.00 &       -3.00 &       -2.00 &       -1.00 &        0.00 &        1.00 &        2.00 &        3.00 &        4.00 \\    
        \hline
        \noalign{\smallskip}
  4.00 &   2.936E+00 &   1.988E+00 &   1.482E+00 &   1.298E+00 &   1.253E+00 &   1.252E+00 &   1.272E+00 &   
  1.312E+00 &  1.356E+00 \\
  4.20 &   2.822E+00 &   1.918E+00 &   1.452E+00 &   1.289E+00 &   1.249E+00 &   1.249E+00 &   1.266E+00 &   
  1.302E+00 &  1.352E+00 \\
  4.40 &   2.713E+00 &   1.852E+00 &   1.425E+00 &   1.281E+00 &   1.247E+00 &   1.246E+00 &   1.261E+00 &   
  1.293E+00 &  1.343E+00 \\
  4.60 &   2.607E+00 &   1.791E+00 &   1.401E+00 &   1.273E+00 &   1.244E+00 &   1.244E+00 &   1.257E+00 &   
  1.285E+00 &  1.333E+00 \\
\hline
        \end{tabular}
\end{table*}

\begin{table*}
        \centering
        \caption{Temperature-averaged free-free Gaunt factor vs. $\gamma^{2}$ for different $u$ and  
        $\kappa-$distribution: $\kappa=5$.}
        \begin{tabular}{rccccccccc}
                \hline\hline
                \noalign{\smallskip}
                log $\gamma^{2}$   & \multicolumn{9}{c}{log $u$} \\
                & -4.00 &       -3.00 &       -2.00 &       -1.00 &        0.00 &        1.00 &        2.00 &        3.00 &        4.00 \\    
                \hline
                \noalign{\smallskip}
 -8.00 &   5.944E+00 &   4.546E+00 &   3.162E+00 &   1.864E+00 &   8.500E-01 &   3.037E-01 &   9.803E-02 &   
 3.107E-02 &  9.828E-03 \\
 -7.80 &   5.944E+00 &   4.546E+00 &   3.162E+00 &   1.864E+00 &   8.501E-01 &   3.037E-01 &   9.804E-02 &   
 3.107E-02 &  9.829E-03 \\
 -7.60 &   5.944E+00 &   4.546E+00 &   3.162E+00 &   1.864E+00 &   8.501E-01 &   3.038E-01 &   9.805E-02 &   
 3.108E-02 &  9.830E-03 \\
 -7.40 &   5.944E+00 &   4.546E+00 &   3.162E+00 &   1.864E+00 &   8.502E-01 &   3.038E-01 &   9.806E-02 &   
 3.108E-02 &  9.832E-03 \\
 -7.20 &   5.944E+00 &   4.546E+00 &   3.162E+00 &   1.864E+00 &   8.503E-01 &   3.038E-01 &   9.808E-02 &   
 3.109E-02 &  9.834E-03 \\
 -7.00 &   5.944E+00 &   4.546E+00 &   3.162E+00 &   1.864E+00 &   8.504E-01 &   3.039E-01 &   9.811E-02 &   
 3.110E-02 &  9.836E-03 \\
\hline
                \end{tabular}
        \end{table*}
        
        \begin{table*}
                \centering
\begin{tabular}{rccccccccc}
        \multicolumn{10}{l}{{\bf{\tablename} \thetable{}} -- Continuation.}\\[0.5ex]
        \hline\hline\\ [-2ex]
        \noalign{\smallskip}
        log $\gamma^{2}$   & \multicolumn{9}{c}{log $u$} \\
        & -4.00 &       -3.00 &       -2.00 &       -1.00 &        0.00 &        1.00 &        2.00 &        3.00 &        4.00 \\    
        \hline
        \noalign{\smallskip}
  4.00 &   2.809E+00 &   1.888E+00 &   1.370E+00 &   1.176E+00 &   1.127E+00 &   1.125E+00 &   1.142E+00 &   
  1.178E+00 &  1.218E+00 \\
  4.20 &   2.701E+00 &   1.817E+00 &   1.339E+00 &   1.166E+00 &   1.124E+00 &   1.122E+00 &   1.137E+00 &   
  1.169E+00 &  1.213E+00 \\
  4.40 &   2.596E+00 &   1.751E+00 &   1.311E+00 &   1.157E+00 &   1.121E+00 &   1.119E+00 &   1.132E+00 &   
  1.161E+00 &  1.206E+00 \\
  4.60 &   2.494E+00 &   1.689E+00 &   1.285E+00 &   1.150E+00 &   1.118E+00 &   1.117E+00 &   1.128E+00 &   
  1.154E+00 &  1.197E+00 \\
\hline
        \end{tabular}
\end{table*}

\begin{table*}
        \centering
        \caption{Temperature-averaged free-free Gaunt factor vs. $\gamma^{2}$ for different $u$ and 
        $\kappa-$distribution: $\kappa=10$.}
        \begin{tabular}{rccccccccc}
                \hline\hline
                \noalign{\smallskip}
                log $\gamma^{2}$   & \multicolumn{9}{c}{log $u$} \\
                & -4.00 &       -3.00 &       -2.00 &       -1.00 &        0.00 &        1.00 &        2.00 &        3.00 &        4.00 \\    
                \hline
                \noalign{\smallskip}
 -8.00 &   5.704E+00 &   4.381E+00 &   3.070E+00 &   1.831E+00 &   8.456E-01 &   3.036E-01 &   9.802E-02 &   
 3.107E-02 &  9.827E-03 \\
 -7.80 &   5.704E+00 &   4.381E+00 &   3.070E+00 &   1.831E+00 &   8.457E-01 &   3.036E-01 &   9.803E-02 &   
 3.107E-02 &  9.828E-03 \\
 -7.60 &   5.704E+00 &   4.381E+00 &   3.070E+00 &   1.831E+00 &   8.457E-01 &   3.036E-01 &   9.804E-02 &   
 3.107E-02 &  9.829E-03 \\
 -7.40 &   5.704E+00 &   4.381E+00 &   3.070E+00 &   1.831E+00 &   8.458E-01 &   3.036E-01 &   9.805E-02 &   
 3.108E-02 &  9.831E-03 \\
 -7.20 &   5.704E+00 &   4.381E+00 &   3.070E+00 &   1.831E+00 &   8.458E-01 &   3.037E-01 &   9.807E-02 &   
 3.109E-02 &  9.832E-03 \\
 -7.00 &   5.704E+00 &   4.381E+00 &   3.070E+00 &   1.831E+00 &   8.459E-01 &   3.037E-01 &   9.809E-02 &   
 3.109E-02 &  9.835E-03 \\
\hline
                \end{tabular}
        \end{table*}
        
\begin{table*}
\centering
\begin{tabular}{rccccccccc}
        \multicolumn{10}{l}{{\bf{\tablename} \thetable{}} -- Continuation.}\\[0.5ex]
        \hline\hline\\ [-2ex]
        \noalign{\smallskip}
        log $\gamma^{2}$   & \multicolumn{9}{c}{log $u$} \\
        & -4.00 &       -3.00 &       -2.00 &       -1.00 &        0.00 &        1.00 &        2.00 &        3.00 &        4.00 \\    
        \hline
        \noalign{\smallskip}
  4.00 &   2.748E+00 &   1.841E+00 &   1.318E+00 &   1.118E+00 &   1.067E+00 &   1.064E+00 &   1.080E+00 &   
  1.114E+00 &  1.152E+00 \\
  4.20 &   2.643E+00 &   1.771E+00 &   1.286E+00 &   1.108E+00 &   1.064E+00 &   1.061E+00 &   1.075E+00 &   
  1.106E+00 &  1.148E+00 \\
  4.40 &   2.540E+00 &   1.704E+00 &   1.257E+00 &   1.099E+00 &   1.061E+00 &   1.059E+00 &   1.071E+00 &   
  1.098E+00 &  1.141E+00 \\
  4.60 &   2.441E+00 &   1.642E+00 &   1.231E+00 &   1.091E+00 &   1.059E+00 &   1.057E+00 &   1.067E+00 &   
  1.091E+00 &  1.132E+00 \\
\hline
        \end{tabular}
\end{table*}

\begin{table*}
        \centering
        \caption{Temperature-averaged free-free Gaunt factor vs. $\gamma^{2}$ for different $u$ and 
        $\kappa-$distribution: $\kappa=15$.}
        \begin{tabular}{rccccccccc}
                \hline\hline
                \noalign{\smallskip}
                log $\gamma^{2}$   & \multicolumn{9}{c}{log $u$} \\
                & -4.00 &       -3.00 &       -2.00 &       -1.00 &        0.00 &        1.00 &        2.00 &        3.00 &        4.00 \\    
                \hline
                \noalign{\smallskip}
 -8.00 &   5.639E+00 &   4.336E+00 &   3.045E+00 &   1.822E+00 &   8.444E-01 &   3.035E-01 &   9.802E-02 &   
 3.107E-02 &  9.827E-03 \\
 -7.80 &   5.639E+00 &   4.336E+00 &   3.045E+00 &   1.822E+00 &   8.445E-01 &   3.035E-01 &   9.802E-02 &   
 3.107E-02 &  9.828E-03 \\
 -7.60 &   5.639E+00 &   4.336E+00 &   3.045E+00 &   1.822E+00 &   8.445E-01 &   3.036E-01 &   9.803E-02 &   
 3.107E-02 &  9.829E-03 \\
 -7.40 &   5.639E+00 &   4.336E+00 &   3.045E+00 &   1.822E+00 &   8.446E-01 &   3.036E-01 &   9.805E-02 &   
 3.108E-02 &  9.830E-03 \\
 -7.20 &   5.639E+00 &   4.336E+00 &   3.045E+00 &   1.822E+00 &   8.446E-01 &   3.036E-01 &   9.806E-02 &   
 3.108E-02 &  9.832E-03 \\
 -7.00 &   5.639E+00 &   4.336E+00 &   3.045E+00 &   1.822E+00 &   8.447E-01 &   3.037E-01 &   9.809E-02 &   
 3.109E-02 &  9.834E-03 \\
\hline
                \end{tabular}
        \end{table*}
        
\begin{table*}
\centering               
\begin{tabular}{rccccccccc}
        \multicolumn{10}{l}{{\bf{\tablename} \thetable{}} -- Continuation.}\\[0.5ex]
        \hline\hline\\ [-2ex]
        \noalign{\smallskip}
        log $\gamma^{2}$   & \multicolumn{9}{c}{log $u$} \\
        & -4.00 &       -3.00 &       -2.00 &       -1.00 &        0.00 &        1.00 &        2.00 &        3.00 &        4.00 \\    
        \hline
        \noalign{\smallskip}
  4.00 &   2.731E+00 &   1.829E+00 &   1.304E+00 &   1.102E+00 &   1.052E+00 &   1.048E+00 &   1.064E+00 &   
  1.097E+00 &  1.134E+00 \\
  4.20 &   2.627E+00 &   1.758E+00 &   1.272E+00 &   1.092E+00 &   1.048E+00 &   1.045E+00 &   1.059E+00 &   
  1.089E+00 &  1.131E+00 \\
  4.40 &   2.525E+00 &   1.692E+00 &   1.242E+00 &   1.083E+00 &   1.045E+00 &   1.043E+00 &   1.055E+00 &   
  1.082E+00 &  1.123E+00 \\
  4.60 &   2.427E+00 &   1.629E+00 &   1.216E+00 &   1.076E+00 &   1.043E+00 &   1.041E+00 &   1.051E+00 &   
  1.075E+00 &  1.115E+00 \\
\hline
        \end{tabular}
\end{table*}

\begin{table*}
        \centering
        \caption{Temperature-averaged free-free Gaunt factor vs. $\gamma^{2}$ for different $u$ and 
        $\kappa-$distribution: $\kappa=25$.}
        \begin{tabular}{rccccccccc}
                \hline\hline
                \noalign{\smallskip}
                log $\gamma^{2}$   & \multicolumn{9}{c}{log $u$} \\
                & -4.00 &       -3.00 &       -2.00 &       -1.00 &        0.00 &        1.00 &        2.00 &        3.00 &        4.00 \\    
                \hline
                \noalign{\smallskip}
 -8.00 &   5.592E+00 &   4.304E+00 &   3.027E+00 &   1.815E+00 &   8.436E-01 &   3.035E-01 &   9.801E-02 &   
 3.107E-02 &  9.827E-03 \\
 -7.80 &   5.592E+00 &   4.304E+00 &   3.027E+00 &   1.815E+00 &   8.436E-01 &   3.035E-01 &   9.802E-02 &   
 3.107E-02 &  9.827E-03 \\
 -7.60 &   5.592E+00 &   4.304E+00 &   3.027E+00 &   1.815E+00 &   8.437E-01 &   3.035E-01 &   9.803E-02 &   
 3.107E-02 &  9.829E-03 \\
 -7.40 &   5.592E+00 &   4.304E+00 &   3.027E+00 &   1.816E+00 &   8.437E-01 &   3.036E-01 &   9.805E-02 &   
 3.108E-02 &  9.830E-03 \\
 -7.20 &   5.592E+00 &   4.304E+00 &   3.027E+00 &   1.816E+00 &   8.438E-01 &   3.036E-01 &   9.806E-02 &   
 3.108E-02 &  9.832E-03 \\
 -7.00 &   5.592E+00 &   4.304E+00 &   3.027E+00 &   1.816E+00 &   8.439E-01 &   3.037E-01 &   9.808E-02 &   
 3.109E-02 &  9.834E-03 \\
\hline
                \end{tabular}
        \end{table*}

\begin{table*}
        \centering
\begin{tabular}{rccccccccc}
        \multicolumn{10}{l}{{\bf{\tablename} \thetable{}} -- Continuation.}\\[0.5ex]
        \hline\hline\\ [-2ex]
        \noalign{\smallskip}
        log $\gamma^{2}$   & \multicolumn{9}{c}{log $u$} \\
        & -4.00 &       -3.00 &       -2.00 &       -1.00 &        0.00 &        1.00 &        2.00 &        3.00 &        4.00 \\    
        \hline
        \noalign{\smallskip}
  4.00 &   2.719E+00 &   1.820E+00 &   1.294E+00 &   1.091E+00 &   1.040E+00 &   1.036E+00 &   1.052E+00 &   
  1.085E+00 &  1.121E+00 \\
  4.20 &   2.616E+00 &   1.749E+00 &   1.261E+00 &   1.081E+00 &   1.037E+00 &   1.034E+00 &   1.047E+00 &   
  1.077E+00 &  1.118E+00 \\
  4.40 &   2.515E+00 &   1.683E+00 &   1.232E+00 &   1.072E+00 &   1.034E+00 &   1.031E+00 &   1.043E+00 &   
  1.069E+00 &  1.111E+00 \\
  4.60 &   2.416E+00 &   1.620E+00 &   1.206E+00 &   1.064E+00 &   1.031E+00 &   1.029E+00 &   1.039E+00 &   
  1.063E+00 &  1.102E+00 \\
\hline
        \end{tabular}
\end{table*}

\begin{table*}
        \centering
        \caption{Temperature-averaged free-free Gaunt factor vs. $\gamma^{2}$ for different $u$ and 
        $\kappa-$distribution: $\kappa=50$.}
        \begin{tabular}{rccccccccc}
                \hline\hline
                \noalign{\smallskip}
                log $\gamma^{2}$   & \multicolumn{9}{c}{log $u$} \\
                & -4.00 &       -3.00 &       -2.00 &       -1.00 &        0.00 &        1.00 &        2.00 &        3.00 &        4.00 \\    
                \hline
                \noalign{\smallskip}
 -8.00 &   5.559E+00 &   4.281E+00 &   3.014E+00 &   1.811E+00 &   8.430E-01 &   3.035E-01 &   9.801E-02 &   
 3.107E-02 &  9.826E-03 \\
 -7.80 &   5.559E+00 &   4.281E+00 &   3.014E+00 &   1.811E+00 &   8.430E-01 &   3.035E-01 &   9.802E-02 &   
 3.107E-02 &  9.827E-03 \\
 -7.60 &   5.559E+00 &   4.281E+00 &   3.014E+00 &   1.811E+00 &   8.431E-01 &   3.035E-01 &   9.803E-02 &   
 3.107E-02 &  9.828E-03 \\
 -7.40 &   5.559E+00 &   4.281E+00 &   3.014E+00 &   1.811E+00 &   8.431E-01 &   3.036E-01 &   9.804E-02 &   
 3.108E-02 &  9.830E-03 \\
 -7.20 &   5.559E+00 &   4.281E+00 &   3.014E+00 &   1.811E+00 &   8.432E-01 &   3.036E-01 &   9.806E-02 &   
 3.108E-02 &  9.832E-03 \\
 -7.00 &   5.559E+00 &   4.281E+00 &   3.014E+00 &   1.811E+00 &   8.433E-01 &   3.036E-01 &   9.808E-02 &   
 3.109E-02 &  9.834E-03 \\
\hline
\end{tabular}
\end{table*}

\begin{table*}
\centering
\begin{tabular}{rccccccccc}
        \multicolumn{10}{l}{{\bf{\tablename} \thetable{}} -- Continuation.}\\[0.5ex]
        \hline\hline\\ [-2ex]
        \noalign{\smallskip}
        log $\gamma^{2}$   & \multicolumn{9}{c}{log $u$} \\
        & -4.00 &       -3.00 &       -2.00 &       -1.00 &        0.00 &        1.00 &        2.00 &        3.00 &        4.00 \\    
        \hline
        \noalign{\smallskip}
           4.00 &   2.711E+00 &   1.814E+00 &   1.287E+00 &   1.083E+00 &   1.032E+00 &   1.028E+00 &   1.043E+00 
           &   1.076E+00 &  1.112E+00 \\
           4.20 &   2.608E+00 &   1.743E+00 &   1.254E+00 &   1.073E+00 &   1.028E+00 &   1.025E+00 &   1.039E+00 
           &   1.068E+00 &  1.109E+00 \\
           4.40 &   2.507E+00 &   1.677E+00 &   1.225E+00 &   1.064E+00 &   1.025E+00 &   1.023E+00 &   1.035E+00 
           &   1.061E+00 &  1.102E+00 \\
           4.60 &   2.409E+00 &   1.614E+00 &   1.198E+00 &   1.056E+00 &   1.023E+00 &   1.021E+00 &   1.031E+00 
           &   1.054E+00 &  1.093E+00 \\
\hline
        \end{tabular}
\end{table*}

\section{Table of the total free-free Gaunt factor for different $\kappa$}

\begin{table*}
\centering
\caption{Total free-free Gaunt factor for the Maxwellian and $\kappa$ distributions of electron energies. 
\label{tot_gff}}
\begin{tabular}{rcccccccc}
\hline\hline
\noalign{\smallskip}
log $\gamma^{2}$   & \multicolumn{8}{c}{$\kappa$} \\
& $\infty$ &    2 &   3 &  5 &  10 &  15 &  25 &   50  \\    
\hline
\noalign{\smallskip}
 -8.00 &  1.103E+00 &   1.221E+00 &   1.152E+00 &   1.125E+00 &   1.112E+00 &   1.109E+00 &   1.106E+00 &   
 1.104E+00 \\
 -7.80 &  1.103E+00 &   1.221E+00 &   1.152E+00 &   1.125E+00 &   1.112E+00 &   1.109E+00 &   1.106E+00 &   
 1.104E+00 \\
 -7.60 &  1.103E+00 &   1.222E+00 &   1.152E+00 &   1.125E+00 &   1.112E+00 &   1.109E+00 &   1.106E+00 &   
 1.104E+00 \\
 -7.40 &  1.103E+00 &   1.222E+00 &   1.152E+00 &   1.126E+00 &   1.112E+00 &   1.109E+00 &   1.106E+00 &   
 1.104E+00 \\
 -7.20 &  1.103E+00 &   1.222E+00 &   1.152E+00 &   1.126E+00 &   1.112E+00 &   1.109E+00 &   1.106E+00 &   
 1.105E+00 \\
 -7.00 &  1.103E+00 &   1.222E+00 &   1.152E+00 &   1.126E+00 &   1.113E+00 &   1.109E+00 &   1.106E+00 &   
 1.105E+00 \\
\hline
\end{tabular}
\end{table*}

\begin{table*}
\centering
\begin{tabular}{rcccccccc}
\multicolumn{9}{l}{{\bf{\tablename} \thetable{}} -- Continuation.}\\[0.5ex]
\hline\hline\\ [-2ex]
\noalign{\smallskip}
log $\gamma^{2}$   & \multicolumn{8}{c}{$\kappa$} \\
& $\infty$ &    2 &   3 &  5 &  10 &  15 &  25 &   50  \\    
\hline
\noalign{\smallskip}
  4.00 &  1.042E+00 &   1.634E+00 &   1.269E+00 &   1.144E+00 &   1.085E+00 &   1.069E+00 &   1.058E+00 &   
  1.050E+00 \\
  4.20 &  1.036E+00 &   1.629E+00 &   1.263E+00 &   1.139E+00 &   1.079E+00 &   1.064E+00 &   1.052E+00 &   
  1.044E+00 \\
  4.40 &  1.031E+00 &   1.624E+00 &   1.259E+00 &   1.134E+00 &   1.075E+00 &   1.059E+00 &   1.047E+00 &   
  1.039E+00 \\
  4.60 &  1.027E+00 &   1.621E+00 &   1.255E+00 &   1.130E+00 &   1.070E+00 &   1.054E+00 &   1.043E+00 &   
  1.035E+00 \\
\hline
\end{tabular}
\end{table*}
\end{appendix}
\end{document}